\documentclass{article}
\usepackage{spconf,amsmath,graphicx}
\usepackage{amssymb}
\usepackage{cite}
\usepackage{color}
\usepackage{epstopdf}
\usepackage{hyperref}       % hyperlinks\textbf{}
\usepackage{url}            % simple URL typesetting
\usepackage{amsmath} 
\usepackage{amsmath} 
\newcommand{\etal}{{\em et al\,.}}       % et al.
\newcommand{\eg}{{\em e.g.}}           % e.g.
\newcommand{\ie}{{\em i.e.}}           % i.e.
\newcommand{\etc}{{\em etc.}}         % etc.
\title{Automatic Data Augmentation via Deep Reinforcement Learning for Effective Kidney Tumor Segmentation}
\name{Author(s) Name(s)\thanks{Thanks to XYZ agency for funding.}}
\address{Author Affiliation(s)}

\name{Tiexin Qin\textsuperscript{\rm 1}, Ziyuan Wang\textsuperscript{\rm 1}, Kelei He\textsuperscript{\rm 2,3}, Yinghuan Shi\textsuperscript{\rm 1,3}\thanks{Corresponding author: Yinghuan Shi (syh@nju.edu.cn)}, Yang Gao\textsuperscript{\rm 1,3}, Dinggang Shen\textsuperscript{\rm 4}
\address{\textsuperscript{\rm 1}National Key Laboratory for Novel Software Technology, Nanjing University, P. R. China \\
\textsuperscript{\rm 2}Medical School of Nanjing University, P. R. China\\
\textsuperscript{\rm 3}National Institute of Healthcare Data Science, Nanjing University, P. R. China\\
\textsuperscript{\rm 4}Department of Radiology and BRIC, UNC Chapel Hill, USA\\}}

\begin{document}
%\ninept
%
\maketitle
\begin{abstract}
\vspace{-0.1cm}

Conventional data augmentation realized by performing simple pre-processing operations (\eg, rotation, crop, \etc) has been validated for its advantage in enhancing the performance for medical image segmentation. However, the data generated by these conventional augmentation methods are random and sometimes harmful to the subsequent segmentation.
In this paper, we developed a novel automatic learning-based data augmentation method for medical image segmentation which models the augmentation task as a trial-and-error procedure using deep reinforcement learning (DRL). In our method, we innovatively combine the data augmentation module and the subsequent segmentation module in an end-to-end training manner with a consistent loss. Specifically, the best sequential combination of different basic operations is automatically learned by directly maximizing the performance improvement (\ie, Dice ratio) on the available validation set.
We extensively evaluated our method on CT kidney tumor segmentation which validated the promising results of our method.
\end{abstract}

\begin{figure*}[htbp]
\centering
\includegraphics[width=0.88\textwidth]{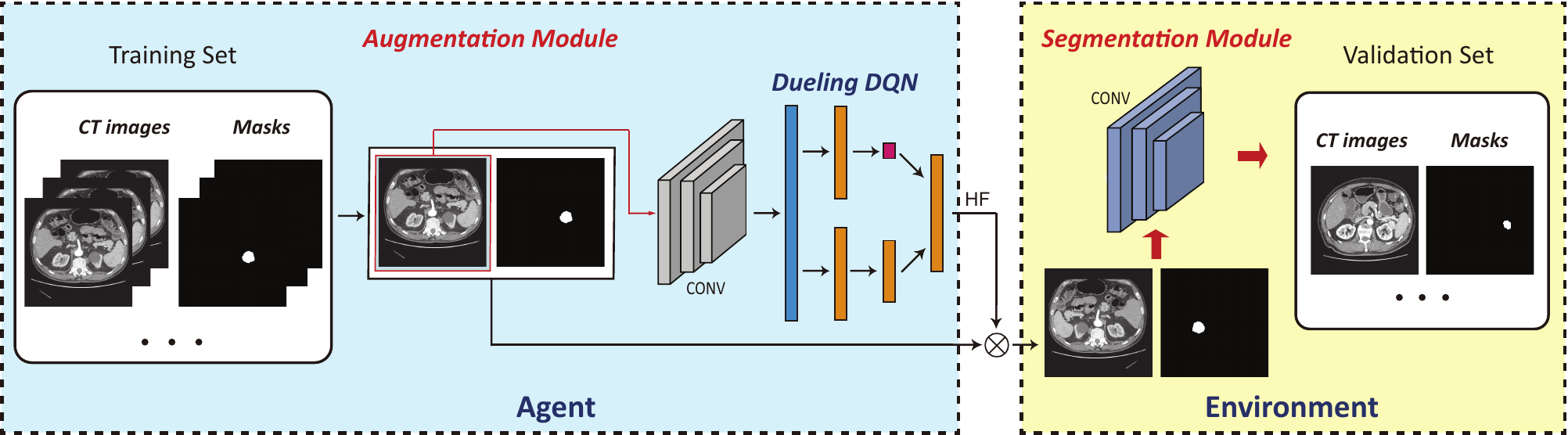}
\caption{The framework of our automatic data augmentation method for medical image segmentation. When randomly selecting a CT image and the corresponding mask from the training set, the initial state $s^0$ is the feature extracted from the CT image. The agent selects and performs an action on current CT image, and the processed image acts as the next input. This process is repeated until the agent decides to terminal this episode.} \label{framework}
\vspace{-0.4cm}
\end{figure*}
\vspace{-0.4cm}
\section{Introduction}
\label{sec:intro}
\vspace{-0.35cm}
%Medical image segmentation is an essential building block of computer aided diagnosis systems. When enough labeled data is available, deep learning based segmentation methods can produce desirable results \cite{Unet2015MICCAI, SegCaps2018arxiv}.
%However, there are less large-scale medical datasets, yet far from sufficient for deep model training, because of:
%1) limited image resource for preserving the patients' privacy
%2) intensive manual effort and high professional knowledge requirement for labeling.
Nowadays, to precisely and automatically delineate the boundary of different organs and tissues, deep learning-based medical image segmentation plays an essential role in computer-aided diagnosis. It is usually believed that when the enough delineated (or labeled) data is available during training, a desirable segmentation result could be expected by using deep learning based methods \cite{Unet2015MICCAI,SegCaps2018arxiv}.
However, as a prerequisite, sufficient training data cannot always be guaranteed in real cases due to
1) limited access permission of medical images for privacy protection, and
2) intensive manual effort and high professional knowledge requirement for manual delineation.
In this case, the performance degeneration usually happens when the trained model adapts to the testing data.

To overcome this challenge from limited data, existing methods intend to increase and diversify the available training images by generating new images from the original ones. This methodology, namely data augmentation, has shown its merits in improving the segmentation accuracy \cite{NIPS2012_4824, Unet2015MICCAI, Ravishankar2017MICCAI, Yang2017MICCAI}.
Typically, data augmentation performs some basic image pre-processing operations (\eg, flipping, cropping and warping) to generate additional training images on the original dataset. The improvement on segmentation tasks is validated by previous studies \cite{PereiraTMI2016,OliveiraIEEE2017}.
However, these methods are highly sensitive to the selection and magnitude of operations \cite{Dosovitskiy2016PAMI} which makes setting the optimal operations an ad-hoc task \cite{Simard2003ICDAR, Ciresan2012CVPR}.

% Related work
Besides, several works attempted to solve the data limitation problem by generating new samples on a simulated or learned data distribution. For example,
%Zhao \etal \cite{Zhao2019CVPR} learn spatial and appearance transform models to synthesize additional labeled training samples based the learned transforms and a single segmented scan.
Zach \etal \cite{Zach2018MIDL} extended Mixup \cite{Zhang2018ICLRmixup} to medical image segmentation by generating virtual data on the marginal region between two classes.
%Liu \etal \cite{GAN2019arxiv} utilized the generative adversarial networks (GANs) to generate realistic images for boosting the performance in semantic segmentation task.
With the same distribution of real and generated virtual data, the generative adversarial networks (GANs) were utilized to generate virtual images for boosting the performance in segmentation task \cite{GAN2019arxiv, neffOAGM2017, ShinSASHIMI2018}.

Unfortunately, previous methods suffer from one major limitation: \emph{the data augmentation module and following segmentation module are trained in a separate way}, where the result of the segmentation module cannot feedback to adjust the augmentation module. Thus, the optimal augmentation method is not guaranteed.

%In some situations, traditional data augmentation like the combination of \emph{e.g.}, \emph{cropping}, \emph{flipping} achieve competitive results than recent methods\cite{Wang2018CVPR}.
To solve the problem, training the augmentation and the segmentation module simultaneously is highly desired. However, the bottleneck of current data augmentation method is that the best operations of augmentation (\ie, to obtain the largest performance improvement) are hard to define. Regarding this challenge, we in this paper present a learning-based automatic data augmentation method for medical image segmentation, which introduces the deep reinforcement learning (DRL) to explore the most effective sequence of image pre-processing operations.
The improvement of the segmentation module may directly influence the policy generated by the data augmentation module. By doing so, the agent can learn specific augmentation policy for each image, with a stable learned operation sequence. Please note that, the proposed method can combine the augmentation module and the subsequent segmentation module in an end-to-end manner.

Our contributions can be summarized in three-fold:
%\begin{itemize}
\textbf{(1)} We model data augmentation for medical image segmentation as a reinforcement learning problem to learn a data-specific augmentation policy.
\textbf{(2)} A joint-learning scheme to integrate a hybrid architecture of Dueling deep Q-learning network (DQN) and a substitute segmentation module is developed where the learned policy directly optimizes the Dice ratio.
\textbf{(3)} Taking the kidney tumor segmentation in CT images as an illustration, the results demonstrate the effectiveness of our method.
%\end{itemize}

\vspace{-0.4cm}
\section{Our Methodology}
\label{sec:Methodology}
\vspace{-0.1cm}

\subsection{Preliminary}
Reinforcement learning (RL) is one of the most representative machine learning paradigm for sequential decision-making problem \cite{Sutton1998}. To continuously interact with the environment, an agent learns the policy to make a sequential decision to maximize an accumulative reward. Mathematically, it can be formulated as a Markov decision process (MDP) $\mathcal{M:=(S,A,P,R,\gamma)}$ where $\mathcal{S}$ is a finite set of states, $\mathcal{A}$ is a finite set of actions that allows the agent to interact with the environment, and $\mathcal{P:S \times A \times S}\to[0,1]$ denotes the probabilities of state transition, respectively. $\mathcal{R:S \times A \times S}\to \mathbb{R}$ denotes the reward function which drives the action of the current agent. $\gamma \in [0,1]$ is the discount factor to control the importance of immediate versus future rewards \cite{Sutton1998}.

During learning, the agent learns from experience to optimize its policy $\mathcal{\pi: S \times A} \to [0,1]$. A widely-used form to represent the optimal policy could be the state-action value function $Q(s,a)$ which is defined as the expected value of the accumulated reward at state $s$ with taking action $a$, $Q(s,a)=E[\sum_{t=0}^{\infty} \gamma \mathcal{R}(s_t,a_t)], s_t\in \mathcal{S}, a_t \in \mathcal{A}$.
This value function could be unrolled recursively by using the Bellman equation, and can thus be solved iteratively as $Q_{i+1}(s,a)=E[r+\gamma\max_{a'} Q(s',a')]$, $r\in \mathcal{R}$.

Recently, Mnih \etal \cite{Mnih2015Nature} proposed an algorithm namely deep Q-learning (DQN) which combines the traditional Q-learning with deep neural network to deal with high dimensional data.
Also, some efforts have been contributed to improve the standard DQN in stability and convergence.
As a representative method, Dueling DQN \cite{DuelingDQN2016ICML} improves origin DQN by integrating two functions: 1) the action-independent value function $V(s)$ to provide an estimate of the value of each state (parameterized by $\sigma$) and 2) an state-dependent action advantage function $C(s,a)$ to calculate the potential benefits of each action.

Formally, the Q-function in Dueling DQN could be defined as follow:
\vspace{-0.2cm}
\begin{displaymath}
\begin{split}
& Q(s^t,a^t;\gamma,\omega,\sigma) = V(s^t;\gamma,\omega,\sigma) \\
& + C(s^{t+1},a^{t+1};\gamma,\omega,\sigma)
-\frac{1}{|C|}\sum_{a^{t+1}}C(s^t,a^{t+1};\gamma,\omega).
\end{split}
\end{displaymath}
\vspace{-0.4cm}

More details could be referred to the original literature \cite{DuelingDQN2016ICML}. Regarding its robust state value estimation and promising performance, we adopt Dueling DQN as the base network for augmentation module.
%\textcolor{red}{(the above paragraphs are not clear!)}
%\fi

\begin{figure*}[htbp]
\centering
\includegraphics[width=5.4in]{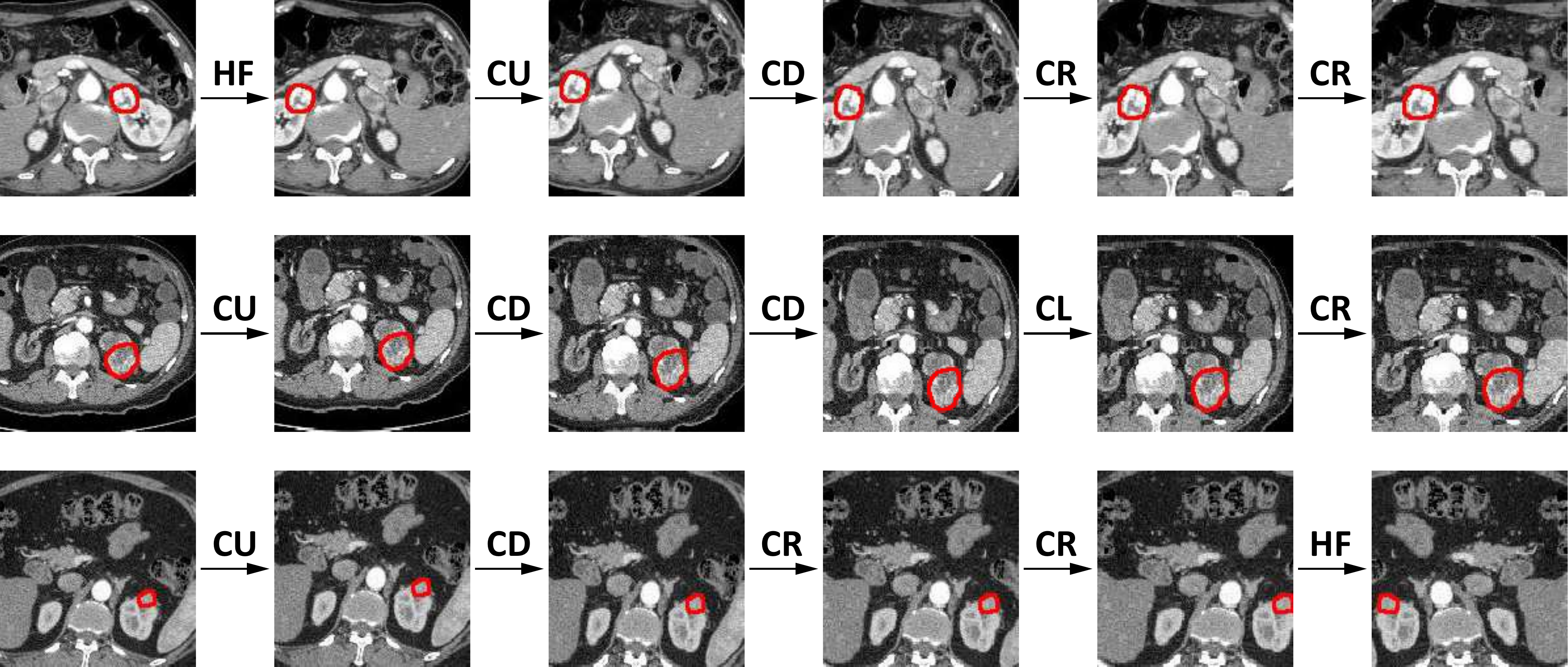}
\vspace{-0.2cm}
\caption{The visualized augmentation steps of some typical cases of our method on a kidney tumor dataset. \textbf{HF} is frequently selected during augmentation since the horizontal flipping operation is widely used for medical image segmentation. Also, crop operations are useful to increase the size of foreground tumors.}
\label{exp2}
\vspace{-0.45cm}
\end{figure*}

\vspace{-0.2cm}
\subsection{Proposed Method}
%\label{sec:method}
% result of augmentation sequence

We now discuss the technical details of our method.
Given a training set $X=\{(x_1,y_1), ...,(x_n,y_n)\}$, where $x_i$ and $y_i$ are the $i$-th CT image and its corresponding segmentation result (\ie, ground truth), respectively, we aim to 1) generate an augmented image $x_i'$ according to $x_i$ which is more effective to improve the segmentation results compared with $x_i$, and at the same time 2) learn a robust segmentation module $f$ trained with original and augmented training set.

As aforementioned, we propose to model joint data augmentation and segmentation as a sequential decision-making problem as illustrated in Fig.~\ref{framework}. Specifically, there are two sub-networks corresponding to the two modules in our method which include 1) a Dueling DQN and 2) a substitute segmentation network. Firstly, the Dueling DQN aims to decide a sequence of basic operations (\ie, actions) to augment the original CT image which could boost the performance of the following segmentation network in a best way which can be measured by the largest performance improvement. Afterwards, the substitute segmentation network is first magnituded by the augmented samples and the performance of the current segmentation network is then evaluated on the validation set for reward calculation.
The components of Dueling DQN are detailed as below.

\vspace{-0.2cm}
\subsubsection{States and Actions}
Regarding that the original CT image ${x_i}$ cannot be directly used as the state due to its high-dimensional representation, we define the state as the high-level feature extracted from ${x_i}$ by a pre-trained segmentation model. This setting is more feasible and reasonable since pre-trained model could provide initial informative features.
Here, we use U-Net \cite{Unet2015MICCAI} as the state extraction network $\phi$ to extract the $t$-step state $s^t$ on $x_i$ as $s^t = \phi(x_i)$.
%Note that, the initial state $s^0$ is the intial feature of this image ${x_i}$.

% Actions may change greatly
%\vspace{-0.3cm}
%\subsubsection{Actions}
During the training procedure, the agent outputs an action $a^t$ according to the current state $s^t$.
We formally define 12 types of basic actions including
\textbf{HF} (horizontally flip the image), \textbf{RT} (rotate the image), \textbf{CL}/\textbf{CR}/\textbf{CU}/\textbf{CD} (crop from left, right, up and down), \textbf{WP} (warp the image), \textbf{ZM} (zoom in the original image into 1.1x), \textbf{AN} (add noise), \textbf{LT} (increase the brightness), \textbf{DK} (decrease the brightness) and \textbf{TM} (terminal this episode).

Specifically, we set the degree of \textbf{RT} to 30. For four crop operations (\ie, \textbf{CL}, \textbf{CR}, \textbf{CU}, \textbf{CD}), the step is set to 20, and the cropped image is resized as before. For \textbf{AN}, the Gaussian noise is added to the normalized image. For \textbf{WP}, the current image is distorted by the PiecewiseAffine operation in imgaug library\footnote{\url{https://imgaug.readthedocs.io/en/latest/index.html}}, For \textbf{LT} and \textbf{DK}, the change magnitude is set to 0.1 (the value of each pixel is normalized in [0,1]).

\vspace{-0.2cm}
\subsubsection{Rewards}
The reward is numerical feedback for each action performed by the agent.
%Normally, the reward function ought to be carefully defined since it could directly influence the final segmentation performance.
With the goal of learning the best augmentation policy, we measure the change of parameters in the segmentation module by the Dice ratio in the validation set $Z$. $Z=\{(z_1,y_1), ...,(z_m,y_m)\}$, where $z_i$ and $y_i$ are the $i$-th CT image and its corresponding segmentation result in the validation set $Z$, respectively, and $m$ is the total number of samples.

Formally, the $t$-step reward $r^t$ is defined as follows:
\begin{equation}
r^t =
\begin{cases}
d^t - d^{t-1},  & \mbox{if }t\mbox{ is not terminal} \\
10(d^t - d^{t-1}), & \mbox{if }t\mbox{ is terminal}
\end{cases}
\end{equation}
where $d^t$ donates the $t$-th step Dice ratio of the segmentation module in validation set defined as below:
\begin{equation}
d^t = \frac{1}{m} \sum_{(z_j,y_j)\in Z} \frac{2|f(\theta_{\mathbf{X}\cup\{x_i^{t}\}};z_j) \cap y_j|}{|f(\theta_{\mathbf{X}\cup\{x_i^{t}\}};z_j)| + |y_j|}
\label{eqn:hybridlossD1}
\end{equation}
where $x_i^t$ denotes the augmented result after $t$-th transformation on the $i$-th image in training set instead of validation set. In Eqn.\,(\ref{eqn:hybridlossD1}), $f(\theta_{\mathbf{X}\cup\{x_i^{t}\}};z_j)$ indicates the segmentation results of $z_j$ by using the model trained on the training set $\mathbf{X}\cup\{x_i^{t}\}$ parameterized by $\theta$.
Typically, $f(\theta_{\mathbf{X}\cup\{x_i^{t}\}};z_j)$ is obtained by fine-tuning $f(\theta_{\mathbf{X}};x_j)$ with $x_i^t$.

\vspace{-0.2cm}
\subsubsection{Agent Learning and Exploiting}
For action learning, we use multiple linear layers to approximate the state-action value function as the common representation of the optimal policy \cite{DuelingDQN2016ICML}. By using the state of each image as input, the corresponding output is the value of each action. The agent chooses the action with the largest calculated value.
%Thanks to this architecture, the learned policy of data augmentation shows its robustness in our evaluation.

For data augmenting, each sample is input into the environment and the agent performs sequential operations on it until it reaches the terminal state. We mix the generated images with original training images as the final augmented set. Thus, for a dataset of size $N$, we generate a dataset of size 2$N$.

\iffalse
\begin{table}[htbp]
\caption{Mean DSC, CD and HD results of different baselines on Kidney tumor dataset.}
\label{tab:acc_kidney}
\centering
\begin{tabular}{ccccc}
\hline
Method & Without Aug. & Traditional Aug. & Random Aug. & Our method \\
\hline
DSC    & 0.758  & 0.766  & 0.770  & \textbf{0.779} \\
CD (mm) & 1.858  & 2.614  & 2.071  & \textbf{1.737} \\
HD (mm) & \textbf{12.710} & 15.685 & 16.371 & 13.653 \\
\hline
\end{tabular}
\end{table}
\fi

%\iffalse
\begin{table*}[htbp]
\caption{The results on kidney tumor dataset.}
\label{tab:acc_kidney}
\renewcommand\arraystretch{1.1}
\centering
%\resizebox{\textwidth}{12mm}{
\small{
\begin{tabular}{cccccccc}
\hline
\textbf{Method} &\textbf{mIoU (\%)} & \textbf{DSC (\%)} & \textbf{PPV (\%)} & \textbf{SEN (\%)} & \textbf{CD (mm)} & \textbf{HD (mm)} & \textbf{ASD (mm)} \\
\hline
Without Aug.     & 65.0     & 74.9   & 67.0  & 90.3	 & 26.31  & 18.00 & 10.14 \\
Traditional Aug. & 74.3     &  83.2  & 79.2  & 92.9    & 15.79  & 13.00 & 5.49 \\
VB-nets \cite{Han2019VBNet}       & 62.2   & 71.8   & 66.9  & 92.5   & 21.10   & 16.55  & 9.58 \\
Neff \etal \cite{neffOAGM2017}    & 47.4   & 57.6   & 51.3  & 88.6   & 53.09   & 28.00 & 17.53 \\
Shin \etal \cite{ShinSASHIMI2018} & 73.3    & 82.0   & 79.6 & 91.8   & 11.67   & 12.08 & 5.42 \\
Costa \etal \cite{CostaTMI2018}   & 53.1   & 62.7   & 56.6 & 91.4   & 34.80  & 22.00 & 13.23 \\
\textbf{Our method} & \textbf{75.8}  & \textbf{84.0} & \textbf{80.2} & \textbf{94.0}  & \textbf{8.86}    & \textbf{8.52} & \textbf{4.01}\\
\hline
\end{tabular}
}
\vspace{-0.3cm}
\end{table*}
%\fi

\vspace{-0.2cm}
\section{Experiments}
\label{sec:experiments}
We now quantitatively and qualitatively report the results of our method. For the dataset, we use the CT kidney tumor dataset collected from a collaborated hospital.
This dataset consists of 68 patients (\ie, subjects) with totally 3,753 CT slices, and there exists one tumor in each slice. The size of each image is $512 \times 512$ and the resolution is $1 \times 1 \text{mm}^2$/pixel.
%We randomly select 500 slices as our dataset to evaluate our method.
%We use this dataset to validate the superiority of our method.

\vspace{-0.2cm}
\subsection{Setting}
% Need to update
%We randomly sample a subset including 500 slices to evaluate our method.
%The dataset is split into training, validation and testing set with a ratio of $5:2:3$.
We randomly partitioned the dataset into training, validation and test subsets, which consist of 50, 5, and 13 different cases, respectively.
The training and validation sets are used to learn the augmentation policy during training Dueling DQN. We set the comparison methods with the same size of augmented set in order for a fair comparison. Besides, the segmentation network structure was kept to be same when comparing with GAN-based augmentation methods. The performance was evaluated by training a segmentation model from scratch on the augmented set.

We used Adam \cite{Kingma2015Adam} optimizer both for Dueling DQN and the segmentation module. For Dueling DQN, the initial learning rate was set to 0.001, the replay buffer was used. For the segmentation module, the initial learning rate was 2.5$\times$10$^{-4}$ and dropped by$(1 - \texttt{epoch} / \texttt{epochs})^{0.9}$. The weight decay was set to $5 \times 10^{-4}$. The learning rate for fine-tuning segmentation module was $5 \times 10^{-5}$. The hype-parameters of all the experiments were set as the same to the proposed method for a fair comparison. We implemented our model with the PyTorch \cite{paszke2017pytorch} Framework on three GTX 2080Ti GPUs.

\vspace{-0.2cm}
\subsection{Results}
\begin{figure}[htbp]
\centering
\includegraphics[width=3.3in]{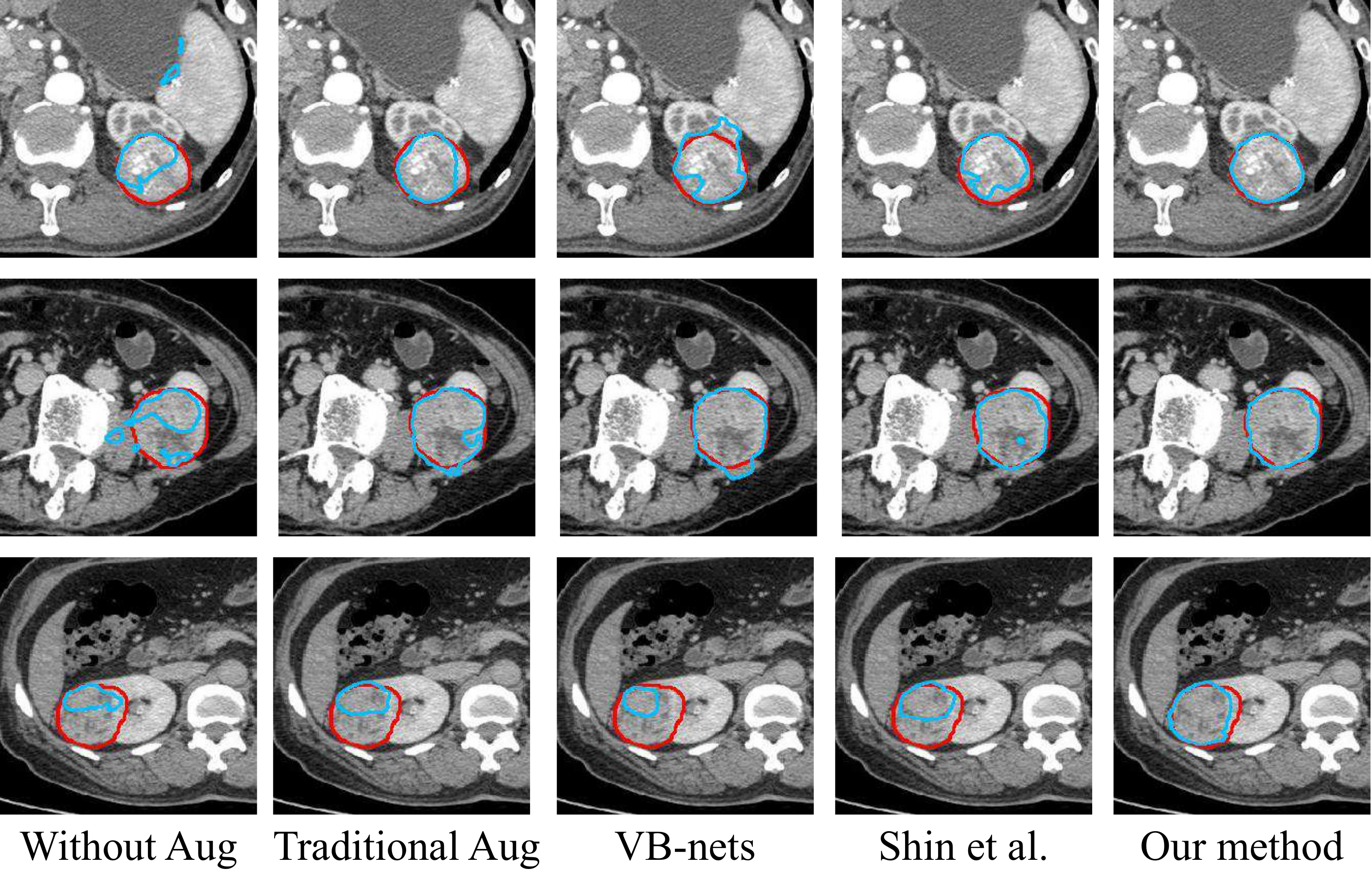}
\caption{The visual segmentation results by training with different augmentation methods. The red contours denote the ground truth provided by the physician and the light blue contours denote the segmentation results of each reported method, respectively. Our method could improve the prediction accuracy especially for the boundary region.}
\label{exp3}
\vspace{-0.2cm}
\end{figure}

We employ the mean Interest of Union (mIoU), Dice ratio (DSC), positive predictive value (PPV), sensitivity(SEN), the median of centroid distance (CD) and Hausdorff distance (HD), and average surface distance (ASD) to evaluate the segmentation accuracy. Table \ref{tab:acc_kidney} reports the results of these baselines. We can infer from the result that our method outperforms other methods in terms of mIoU, DSC, SEN, CD, HD and ASD. Also, we show some typical augmentation examples in Figure \ref{exp2}. It can be observed that \textbf{HF} is frequently selected during augmentation since flipping the image in a horizontal direction is widely used for medical image segmentation. Also, crop operations are useful to increase the size of foreground tumors.
In addition, several segmentation results are reported in Figure \ref{exp3}, showing that our method is able to localize the tumor boundary more accurately than that of other baselines.

%Since the training set is too small for U-Net training, the prediction may fail with full black. In these cases, we substitute result for the worst.

\iffalse
\begin{table}[htbp]
\caption{The results on kidney tumor dataset. The best results are in bold.}
\label{tab:acc_kidney}
\renewcommand\arraystretch{1.2}
\centering
%\resizebox{\textwidth}{12mm}{
\footnotesize{
\begin{tabular}{cccccc}
\hline
\textbf{Method} &\textbf{mIoU} & \textbf{DSC} & \textbf{SEN} & \textbf{CD(mm)} & \textbf{HD(mm)}\\
\hline
Without Aug.     & 0.6502     & 0.7485   & 0.9027	& 584.081  & 18.000 \\
Traditional Aug. & 0.7428     &  0.8324  & 0.9287  & 208.434  & 13.000 \\
VB-nets \cite{Han2019VBNet}       & 0.6221   & 0.7180   & 0.9253   & 242.079   & 16.553\\
Neff \etal \cite{neffOAGM2017}    & 0.4737   & 0.5756   & 0.8864   & 678.746   & 28.000\\
Shin \etal \cite{ShinSASHIMI2018} & 0.733    & 0.8196   & 0.9175   & \textbf{107.940}   & 13.038\\
Costa \etal \cite{CostaTMI2018}   & 0.5307   & 0.6265   & 0.9138   & 1067.747  & 22.000 \\
\textbf{Our method} & \textbf{0.7578}  & \textbf{0.8399} & \textbf{0.9404}  & 132.095    & \textbf{8.515} \\
\hline
\end{tabular}
}
\end{table}
\fi

% result of segmentation

\vspace{-0.2cm}
\section{Conclusion}
\label{sec:conclusion}
We propose a learning-based augmentation method to deal with the case of insufficient labeled images in CT kidney tumor segmentation. We model data augmentation as a sequential decision making problem, and utilize the DRL to search for the most effective sequence to augment each image automatically. Therefore, the result of the substitute segmentation module could return the feedback to adjust the augmentation method. To our knowledge, this is the first attempt to combine data augmentation and segmentation in an end-to-end manner. The results demonstrated the effectiveness of our method by comparing with the previous augmentation methods.

% \iffalse
Our future directions include: 1) extending to other medical image segmentation tasks to validate the generalization of the proposed method \cite{ShiTPAMI2015, YuTIP2019}, 2) combining with GAN-based methods \cite{ShinSASHIMI2018, CostaTMI2018} to increase the diversity of generated data.
% \fi

\vspace{-0.3cm}
\subsection*{Acknowledgments}
\label{sec:acknowledgments}
The work was supported by the National Key Research and Development Program of China (2019YFC0118300), NSFC (61432008, 61673203, 81927808), and Jiangsu Provincial Key Research and Development Project (BE2018610).

\vfill\pagebreak

\bibliographystyle{IEEEbib}
\bibliography{myref}

\end{document}